\documentclass[aps,prl,twocolumn,superscriptaddress]{revtex4-1}

\usepackage{graphicx}
\usepackage{xcolor}
\usepackage{soul}

\newcommand{\beq}{\begin{equation}}
\newcommand{\eeq}{\end{equation}}

\begin{document}

\title{An ultrastable silicon cavity in a continuously operating closed-cycle cryostat at 4 K}
\author{W. Zhang}
\email[]{w.zhang@jila.colorado.edu }
\author{J. M. Robinson}
\author{L. Sonderhouse}
\author{E. Oelker}
\author{C. Benko}
\author{J. L. Hall}
\affiliation{JILA, NIST and University of Colorado, 440 UCB, Boulder, Colorado 80309, USA}
\author{T. Legero}
\author{D. G. Matei}
\author{F. Riehle}
\author{U. Sterr}
\affiliation{Physikalisch-Technische Bundesanstalt, Bundesallee 100, 38116 Braunschweig, Germany}
\author{J. Ye}
\affiliation{JILA, NIST and University of Colorado, 440 UCB, Boulder, Colorado 80309, USA}

\date{\today}

\begin{abstract}
We report on a laser locked to a silicon cavity operating continuously at 4~K with $1 \times 10^{-16}$ instability and a median linewidth of 17~mHz at 1542~nm. This is a ten-fold improvement in short-term instability, and a $10^4$ improvement in linewidth, over previous sub-10 K systems. Operating at low temperatures reduces the thermal noise floor, and thus is advantageous toward reaching an instability of $10^{-18}$, a long-sought goal of the optical clock community. The performance of this system demonstrates the technical readiness for the development of the next generation of ultrastable lasers that operate with ultranarrow linewidth and long-term stability without user intervention.
\end{abstract}
\maketitle
\begin{figure}[t]
\centerline{\includegraphics[scale=0.38]{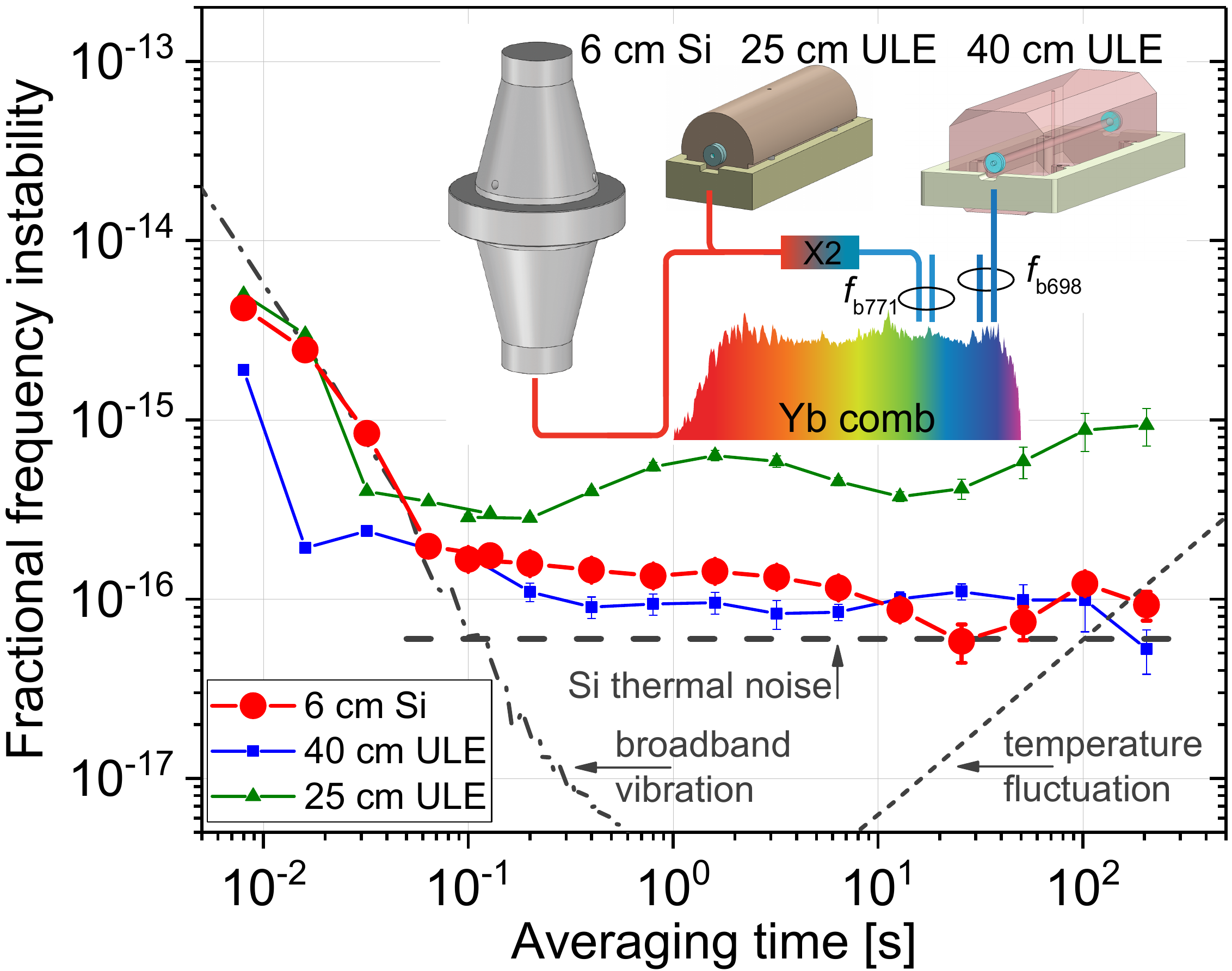}}
\caption{Instability of the 6~cm silicon cavity (circles), 40~cm ULE cavity (squares), and 25~cm ULE cavity (triangles), derived from a three-cornered-hat measurement, whose setup is shown in the inset. Between 0.06 and 100~s, the silicon cavity has an instability around \mbox{$1\times 10^{-16}$}.}
\label{fig:3}
\end{figure}

Ultrastable lasers are a vital component of precision measurement science. Several applications require lasers with high phase coherence, such as laser interferometric gravitational wave detectors~\cite{Rana2014}, very-long-baseline interferometers for radio astronomy~\cite{VLBI}, and low phase-noise microwave synthesizers~\cite{Fortier2011}. Recent advances in laser stabilization have powered the development of the optical atomic clock, which can now measure the passage of time with record accuracy and precision~\cite{Bloom2014,Hinkley2013}. Record atom-light coherence is currently limited by the instability of the clock laser~\cite{campbell2017fermi}. In principle, these clocks can reach an instability limited only by their few mHz linewidth clock transitions.  However, this will require a new generation of ultrastable lasers with fractional frequency stabilities at the $10^{-18}$ level.  Further improvements in clock performance will pay dividends, opening up new applications for optical clocks in dark matter detection~\cite{Derevianko2014,StadnikPRA2016}, quantum many-body physics~\cite{Martin2013}, and gravitational wave astronomy~\cite{Kolkowitz2016}.\\
\indent Crystalline silicon has emerged as a promising material for the next generation of ultrastable optical cavities~\cite{Kessler2012,Hagemann2014}. Recently, a laser stabilized to a silicon cavity operated at 124~K demonstrated a record instability of $4\times 10^{-17}$, limited by thermomechanical noise in the optical coatings~\cite{Matei2017}. Reducing the operating temperature to 4~K can offer several advantages. First, coating thermal noise is greatly reduced at 4~K, bringing $10^{-18}$ instability within reach. Second, both the value and slope of the coefficient of thermal expansion (CTE) decrease rapidly near absolute zero, which greatly relaxes the temperature stability requirements. Closed-cycle cryocoolers allow for maintenance-free operation without cryogen refilling. Ultrastable optical cavities in closed-cycle cryostats are suitable for space-based missions such as gravitational wave detection using atomic clocks~\cite{Kolkowitz2016}. Low-temperature optical cavities have also achieved ultralow drift, making them useful as standalone local oscillators~\cite{Hagemann2014}, and opening up applications in testing Lorentz invariance~\cite{MullerLorentzPRL2003,BraxmaierRelativityPRL2001} and Einstein's equivalence principle~\cite{Wiens2016}. \\
\indent Achieving a low short-term instability with a closed-cycle cryocooler at 4~K presents two primary challenges. The closed-cycle cryocooler generates significant vibrations. Furthermore, constructing a thermal filter to suppress the temperature fluctuations of the cryocooler is difficult at low temperatures because the specific heat of most materials vanishes with decreasing temperature.\\
\indent In this Letter, we present the first continuous operation of a sub-10 K ultrastable cavity with a median laser linewidth of 17~mHz at 1542 nm and $1\times10^{-16}$ instability. This is a $10^4$ improvement in linewidth and a ten-fold improvement in short-term instability over the previous best sub-10~K optical cavity~\cite{Wiens2016,Seel1997}. These results are attained using a custom-designed cryostat that suppresses temperature fluctuations and vibration noise to below the thermal noise floor of $6 \times 10^{-17}$, which is computed taking into account the operating temperature, the cavity length, and the beam spot size~\cite{Cole2013}. With further upgrades on cavity length and mirror coatings, we now have a clear path forward for a continuous operating optical local oscillator below the $10^{-17}$ level instability.\\
\indent Our system comprises a 6~cm-long silicon cavity with 1~m radius of curvature silicon substrates and SiO$_2$/Ta$_2$O$_5$ dielectric coatings, providing a finesse of 400,000 for the TEM00 mode. The cavity operates at 4~K using a two-stage Gifford-McMahon  closed-cycle cryocooler~\cite{Montana}. The temperature fluctuations of the cryocooler are typically $20~\mathrm{mK}$ at an averaging time of 6~s. Given a CTE of $\approx 2 \times 10^{-11}$/K at 4~K, the corresponding frequency instability is on the $10^{-13}$ level, which is several orders of magnitude above the thermal noise floor.\\
\begin{figure}[t]
\centering
\includegraphics[scale=.5]{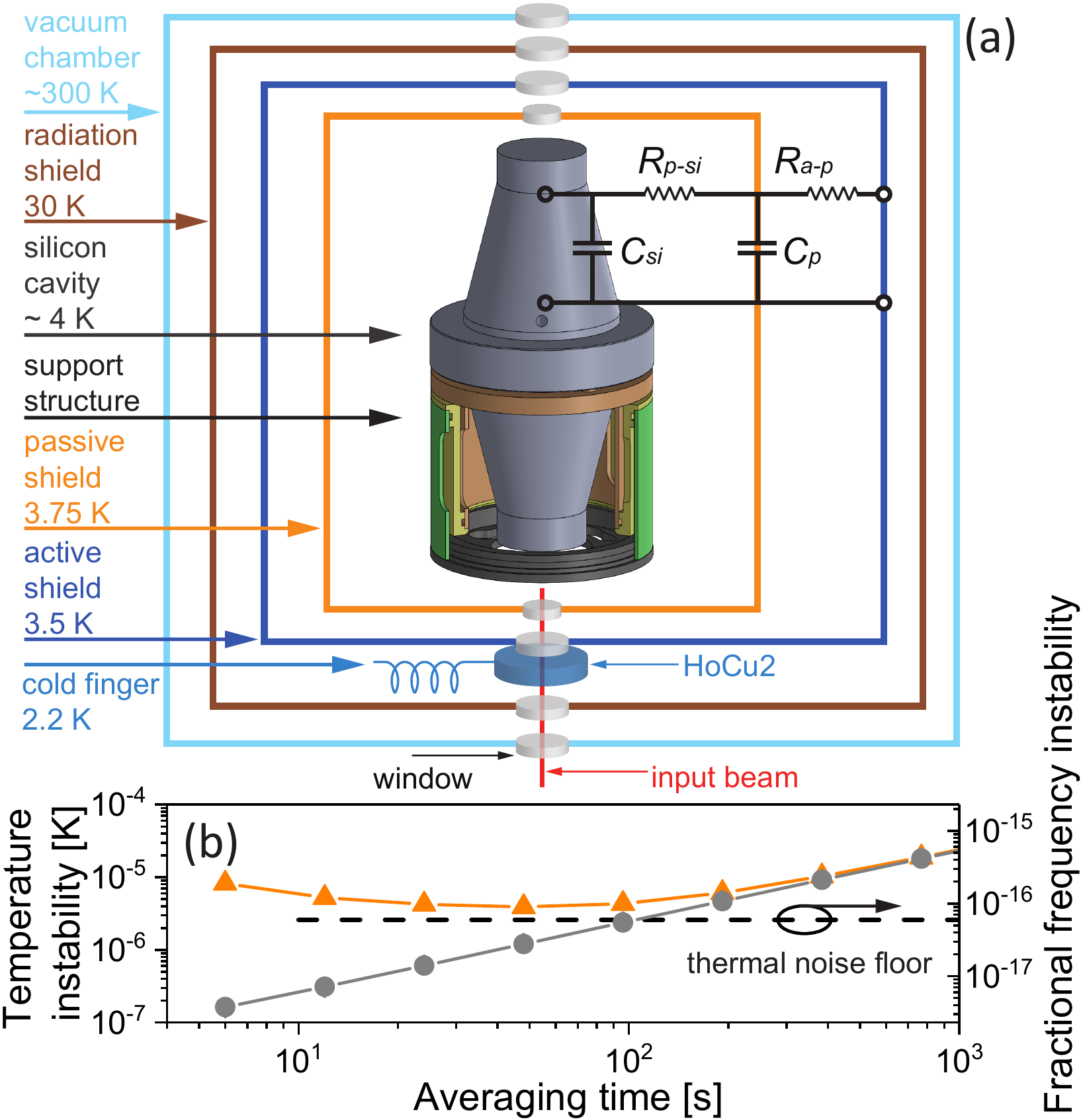}
\caption{(a) Schematic of the chamber. The second stage of the cryocooler (not shown) is coupled to the cold finger. The thermal model is equivalent to an electrical second-order low-pass filter. The passive shield heat capacity is $C_p= 3.5$ J/K, and the thermal resistivity between the active and passive shields is $R_{\text{a-p}}= 395$ K/W. The thermal resistivity between the passive shield and the cavity is $R_{\text{p-Si}}= 2.3\times 10^5$ K/W, and the cavity heat capacity is $C_{\text{Si}}= 2.6$ mJ/K. A section of the cavity's support structure is also shown. (b) Measured temperature fluctuations of the passive shield (triangles), which are used to estimate temperature fluctuations on the cavity (circles). }
\label{fig:1}
\end{figure}
\indent To suppress temperature fluctuations from the cryocooler, we use a custom, multi-stage thermal damping system (Fig.~\ref{fig:1}a). The shields are designed to maximize their heat capacity, while the connections between the shields have a low thermal conductance. An aluminum radiation shield is cooled to 30~K by thermally coupling it to the first stage of the cryocooler. The cooling power for the next shield, called the active shield (made of copper), comes from the second stage of the cryocooler (2.2~K) through a mechanically flexible connection called the cold finger. Between the cold finger and the bottom of the active shield, a cylindrical plate made of holmium copper (HoCu$_2$), which has a specific heat $2500$ times larger than copper at 4~K, is inserted to passively suppress temperature fluctuations from the cold finger. The temperature of the active shield is controlled using a resistive heater. For additional thermal damping, we insert a passive shield between the active shield and the cavity, which is mounted on a cylindrical ring made of G10 fiber glass (not shown). The low thermal conductivity of G10 and the ring's small cross-sectional area ($96\,\mathrm{mm}^2$) minimize the thermal conductance between the two shields. The bottom of the passive shield is constructed out of HoCu$_2$. A cup-shaped structure made of G10 is attached to the base of the passive shield, upon which the cavity is supported. The support structure has three nested layers to increase its effective thermal path, and has three 1~mm-long G10 rods on top to support the cavity. The contact area of each rod is 0.75~mm$^2$ to minimize the thermal conductivity. The thermal model is equivalent to an electrical second-order resistor-capacitor low-pass filter as shown in Fig.~\ref{fig:1}a. The time constant is $\approx 1400$~s between the active and passive shields, and is $\approx 600$~s between the passive shield and the cavity.\\
\begin{figure}
\centerline{\includegraphics[scale=.37]{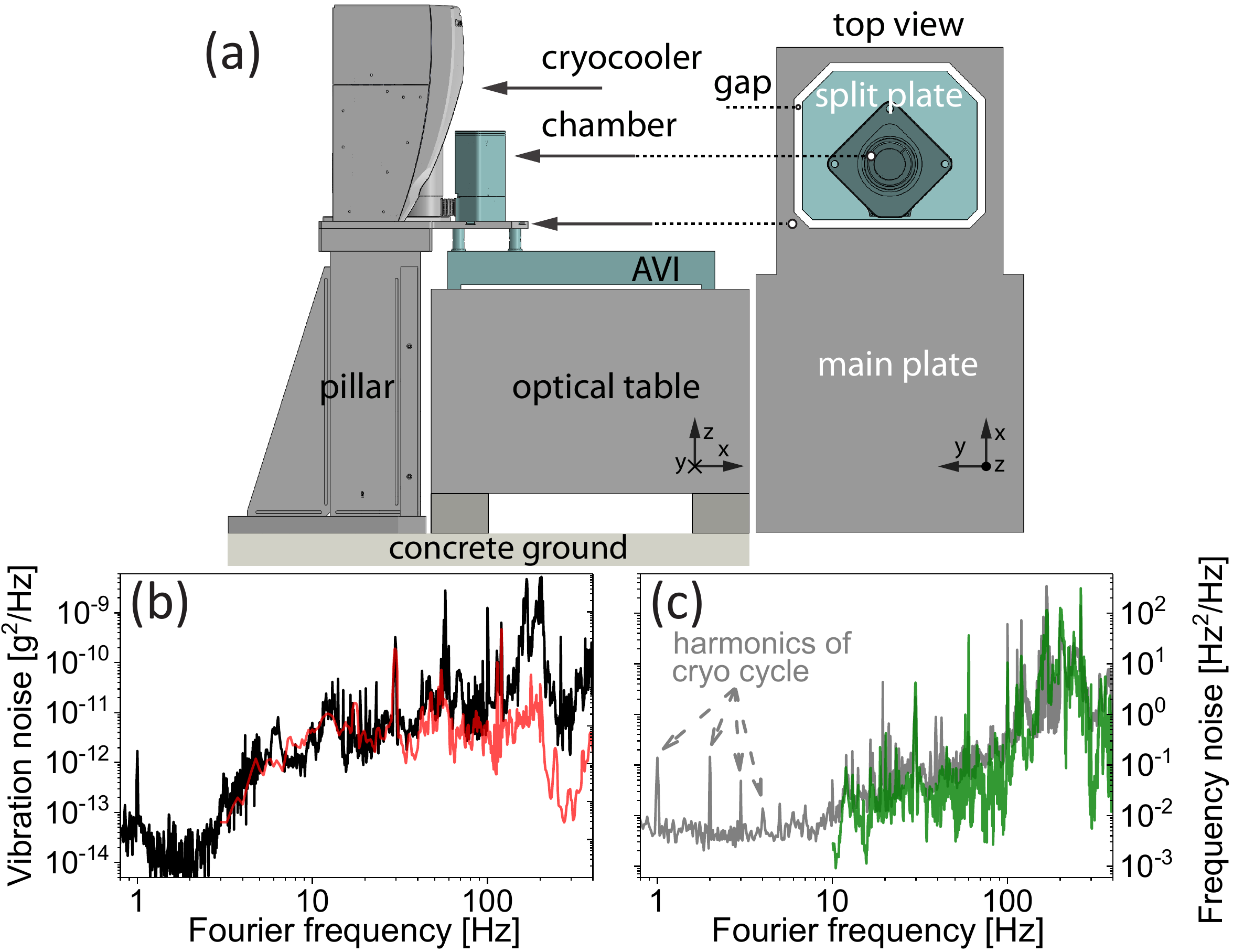}}
\caption{(a) Mounting of the cryostat and a top-view of the separated platform, consisting of the main plate (gray) that the cryocooler sits on and the split plate (light cyan) that supports the vacuum chamber (dark cyan). The split plate is mechanically decoupled from the main plate, and is directly attached to the active vibration isolation (AVI) table. (b) Vibration noise on top of the AVI table when the cryostat is on (black) and off (red). (c) Frequency noise of the beat between the silicon cavity and the 40~cm ULE cavity (gray), and the vibration-induced frequency noise (green), calculated from the measured vibration noise in the region of good spectral coherence above 10 Hz.}
\label{fig:2}
\end{figure}
\indent To assess the performance of the multi-stage thermal damping system, we measure the stability of the various shields and the corresponding transfer functions. The active shield is stabilized to 3.5~K with 0.5~mK instability at an averaging time of $\tau=6$~s, and averages down as $\tau^{-1}$. The temperature of the silicon cavity is estimated to be $T_{\text{Si}} \approx 4$~K. To estimate the temperature fluctuations of the cavity, we convolve the passive shield fluctuations (orange triangles in Fig~\ref{fig:1}b) with the measured transfer function from the passive shield to the cavity. The corresponding temperature-induced frequency fluctuations (gray circles in Fig~\ref{fig:1}b) are below the expected thermal noise floor for averaging times up to 100~s. Fluctuations from room temperature contribute to the rise in instability at long averaging times.  We expect additional thermal control of the cavity enclosure to improve the long-term stability.\\
\indent Vibrations are a significant technical noise source for sub-10~K cavities~\cite{Wiens2016,Seel1997}. 
We address the vibration-induced frequency noise by designing both the cavity and the mechanical layout as shown in Fig \ref{fig:2}a.\\    
\indent Vibration sensitivity typically scales with cavity length~\cite{notcutt2005simple}. Consequently, we design a cavity that is only 6~cm long using finite element analysis (FEA)~\cite{ChenPhysRevA2006}. This 6~cm cavity allows for a compact support structure with high mechanical resonance frequencies. The lowest FEA-simulated mechanical resonance of the cavity with the support structure is 700~Hz. However, in terms of fractional instability, a shorter cavity has a higher thermal noise floor being inversely proportional to the cavity length, and the susceptibility to other technical noise sources like residual amplitude modulation (RAM) increases. The cavity is wider in the middle with a diameter of 45~mm, and tapers down to 16~mm at each end. The cavity is vertically mounted and has a ring 0.5~mm higher than its mid plane, where a support structure provides mechanical contact between the cavity and the chamber. The vertical axis of the cavity is aligned with the $\langle111\rangle$ direction of silicon. The cavity is supported at 3 points in the (111) plane, where the elasticity is anisotropic with a periodicity of 120$^{\circ}$. The vertical vibration sensitivity is a function of the angle between the cavity and the support structure~\cite{Matei2016Postdam}. We aim to support the cavity along the $\langle10$\,-$1\rangle$ direction, which gives a simulated vertical vibration sensitivity of $2.5 \times 10^{-12}$/g/degree, where the angle corresponds to the azimuthal angle between $\langle10$\,-$1\rangle$ and the tripod support. The simulated horizontal sensitivity is $4.3 \times 10^{-11}$/g/mm, where the distance refers to the transverse displacement of the beam from the cavity axis. The vibrations on top of the cryocooler are measured to be $\approx 0.1$~g (g=9.8~m/s$^2$) peak-to-peak. We would expect frequency deviations of hundreds of Hz given a transverse displacement of 0.3 mm and a 3 degree rotation from the optimal support position. Thus, the design of the mechanical layout must provide a vibration damping factor of 4 to 5 orders of magnitude. \\
\indent  The mechanical design (Fig.~\ref{fig:2}a) isolates the cavity from the vibrations of the cryocooler. The cryocooler sits on a main plate that is firmly fixed to a monolithic aluminum pillar that is bolted to the concrete foundation of the laboratory. There are three mandatory mechanical links between the cryocooler and the chamber: a cold finger to cool the active shield, a thermal link connecting the first stage of the cryocooler to the 30~K shield, and a vacuum bellows to isolate them from the environment. All other mechanical connections are eliminated by placing the vacuum chamber on a split plate, which is mechanically separated from the main plate (see the top view in Fig.~\ref{fig:2}a). The gap between the main plate and the split plate is 1.5~mm, and is exaggerated in the figure. The split plate is supported by three rods that are bolted onto an active vibration isolation (AVI) table. The AVI rests on an optical table that is not floating.\\
\indent To characterize the vibration isolation, we place accelerometers on the AVI table and on top of the cryocooler to measure the vibration noise along the x, y, and z directions. The black line in Fig.~\ref{fig:2}b shows the quadrature sum of the three directions on the AVI table. We find that the vibrations on the AVI table are lower than on top of the cryocooler by 4 to 5 orders of magnitude between 1~Hz and 10~Hz, and by 3 orders of magnitude between 10~Hz and 400~Hz. The red line in Fig.~\ref{fig:2}b shows the vibration noise when the cryocooler is off. The spectra are nearly identical up to 100~Hz, though the cryocooler contributes some additional noise at higher frequencies. \\
\indent We simultaneously measure the vibration noise and a heterodyne beat between the silicon cavity and a stable laser at 698~nm based on a 40~cm-long ultra-low expansion glass (ULE) cavity~\cite{Bishof2013} using a Yb frequency comb~\cite{Benko2012}. To quantify the correlation between vibrations measured on the AVI table and the frequency noise of the beat note we introduce the~\textit{magnitude-squared coherence}:
$C_{a_i \nu}(f) = |S_{a_i \nu}(f)|^2/[S_{\nu \nu}(f)S_{a_i a_i}(f)]$. Here, $S_{a_i \nu}(f)$ is the cross-spectral density between frequency fluctuations of the beat signal and acceleration noise in the $i$th direction ($i = x,y,z$), and $S_{\nu \nu}(f)$ and $S_{a_i a_i}(f)$ are the autospectral densities for frequency noise and accelerations, respectively.  A nonzero coherence value at a given $f$ indicates that the laser noise and vibrations are correlated at that spectral frequency.  This is an important prerequisite for measuring transfer functions from accelerations to frequency noise, $H_{a_i}(f) = S_{a_i \nu}(f)/S_{a_i a_i}(f)$, since phase estimates in the cross spectrum are only trustworthy where significant frequency-domain correlation exists.  From 10~Hz to 400~Hz, we measure appreciable coherence between frequency noise and vibrations along all three directions. Within this frequency range, we compute transfer functions from each accelerometer on the AVI table to the beat frequency spectrum by averaging $H_{a_i}(f)$ over an hour-long dataset.  By multiplying each acceleration noise spectrum by the appropriate transfer function and summing the three directions in quadrature, we compute an estimate of the vibration-induced frequency noise in this band. In Fig.~\ref{fig:2}c, this estimate (green) is plotted along with the measured frequency noise spectrum (gray). We have assumed uncorrelated vibrations in the three directions, and a more complete analysis would take into account finite correlations between the different directions. \\
\indent The sharp spikes between 1 and 10~Hz are observed at frequencies corresponding to harmonics of the 1~Hz cryocooler cycle rate. We measure near unity coherence between the vibrations on top of the cryocooler and the frequency noise at these discrete frequencies. The frequency noise spectrum between 1 and 10~Hz is primarily limited by a white noise floor, which does not show coherence with the vibrations of the cryocooler, and could be due to nonlinear vibrational coupling or unsuppressed RAM.\\ 
\indent Fig.~\ref{fig:3} shows three dominant noise sources, vibrations (dash-dotted line), temperature fluctuations (dotted line), and the expected thermal noise floor of the silicon cavity (dashed line), equal to $6 \times 10^{-17}$. The vibration-induced instability is computed from the measured broadband vibration noise shown in Fig.~\ref{fig:2}c. The temperature-induced instability is equivalent to the gray line in Fig.~\ref{fig:1}b. All additional noise sources must also be reduced to below the expected thermal noise floor. The RAM with active stabilization~\cite{Wong85,Zhang2014}, uncompensated fiber noise, servo electronics noise, and the stabilized laser intensity noise are below the $2 \times 10^{-17}$ level and are not shown here. The sum of these additional technical noise sources is around $3 \times 10^{-17}$ for $0.3\, \text{s}<\tau<100\,\text{s}$. \\
\indent To investigate the instability of the ultrastable laser system, we perform a three-cornered-hat (TCH) comparison with two other ultrastable lasers (Fig.~\ref{fig:3} inset). One reference laser is locked to a 25~cm ULE cavity at 1542~nm, while the second is locked to a 40~cm ULE cavity at 698~nm~\cite{Bishof2013}. Each 1542~nm laser is frequency doubled to 771~nm and compared with a Yb comb phase-locked to the 40~cm ULE cavity.  The heterodyne beat between the 25~cm ULE cavity and the silicon cavity is directly measured at 1542~nm.  We measure the beats synchronously using a dead-time-free counter in lambda mode. We bandpass each beat signal with a bandwidth of 1~kHz in order to approximate the correct sensitivity function. We use fourteen data sets that are 40~s long for $8\,  \text{ms}<\tau<0.1 \,\text{s}$, and nine data sets that are 1000~s long to compute $0.1 \,\text{s}<\tau<200\,  \text{s}$. Each dataset has an independent linear drift removed, and we compute the modified Allan deviation for each laser using a correlated TCH analysis~\cite{TCH}.  We compute a statistical average of each individual laser stability, with error bars that represent the standard error of the mean. The red circles in Fig.~\ref{fig:3} show that the silicon cavity supports a low instability around $1 \times 10^{-16}$, which is close to the expected thermal noise floor over several decades of averaging times. For $\tau<0.06\,\text{s}$, the instability is limited by vibrations. \\
\begin{figure}[t]
\centerline{\includegraphics[scale=.35]{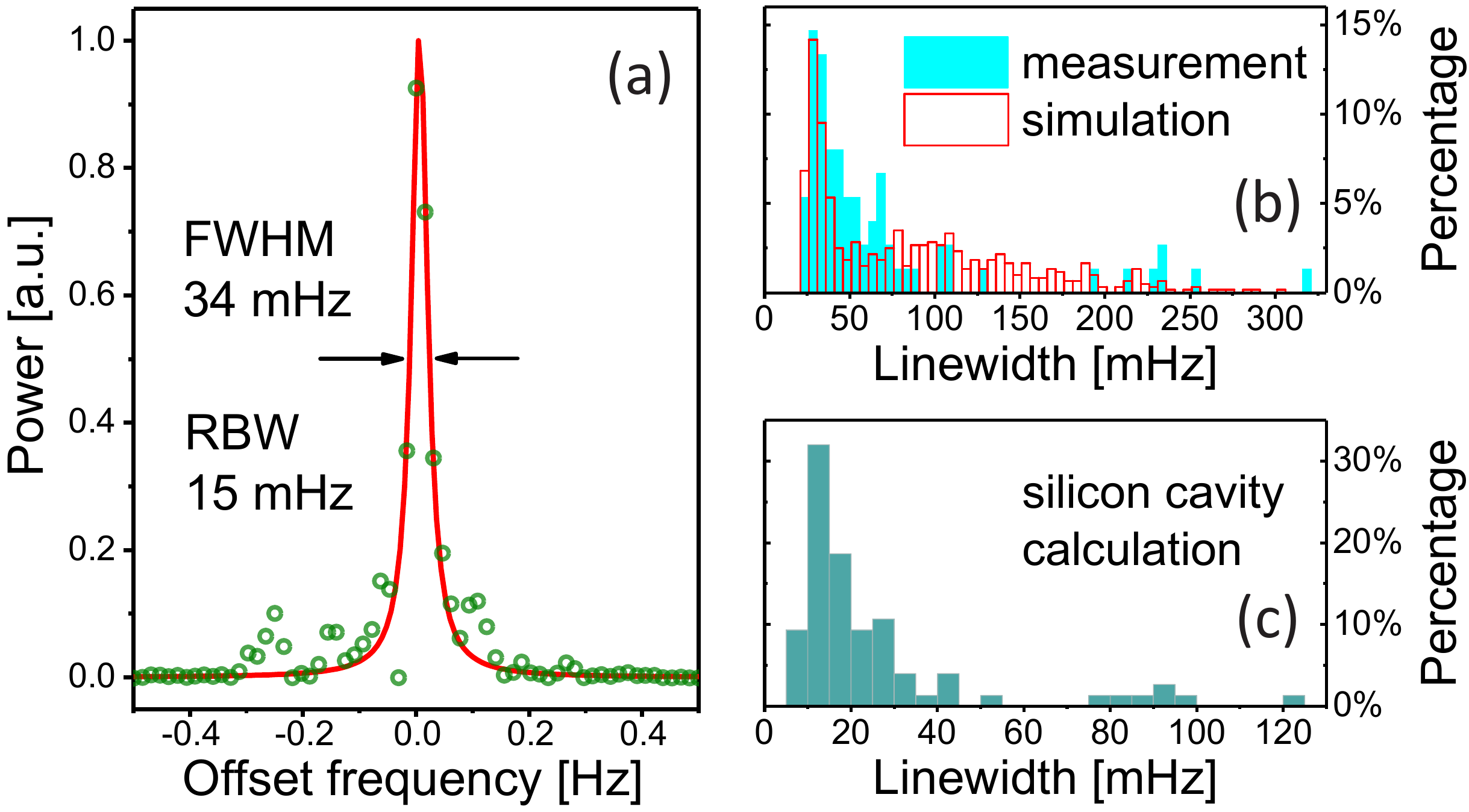}}
\caption{ (a) Beat note at 771~nm (green circles), fit to a Lorentzian (red line). (b) Histogram of the beat note linewidths for 75 measurements (blue) and simulations (red) at 771~nm. (c) Computed laser linewidth of the silicon cavity at 1542 nm. The median of the distribution is 17~mHz.}
\label{fig:4}
\end{figure}
\indent The linewidth of the laser locked to the silicon cavity is extracted from an optical beat at 771~nm between the silicon cavity and the 40~cm ULE cavity through the Yb comb. After removing a linear drift of 20~mHz/s dominated by the 40~cm ULE cavity, the beat is down converted to 10~kHz and recorded using a fast Fourier transform analyzer with a resolution bandwidth of 15~mHz and a measurement time of 64~s. We make 75 measurements of the heterodyne beat. One example, shown in Fig.~\ref{fig:4}a, yields a fitted Lorentz full-width half-maximum (FWHM) of 34 mHz. To simulate the linewidth distribution~\cite{Matei2017}, we generate time-domain data by assuming the silicon-ULE (40~cm) beat has a flicker floor of $1.4 \times 10^{-16}$, which is the quadrature sum of the silicon cavity's instability of $\sigma_{Si}=1.1\times 10^{-16}$ and the ULE cavity's instability of $\sigma_{ULE}=8.8\times 10^{-17}$. Fig.~\ref{fig:4}b shows a reasonable agreement between the simulated and measured beat note. We multiply the measured linewidth distribution by $\sigma_{Si}/\sqrt{\sigma_{Si}^2 + \sigma_{ULE}^2}$ to calculate the laser linewidth distribution of the silicon cavity. Fig.~\ref{fig:4}c shows the computed results, which reveal that the median of the linewidth distribution is 17~mHz at 1542~nm, a $10^4$ improvement over previously reported low-temperature cryogenic cavities~\cite{Seel1997}.\\ 
\indent The drift of the silicon cavity is extracted by locking it to a Sr optical lattice clock~\cite{campbell2017fermi} through the Yb comb. Over a three-day period, the linear drift was \mbox{$<$ 1 mHz/s} at 1542~nm ($5 \times  10^{-18}/$s). Further investigation of the cavity's long-term drift is ongoing.\\
\indent We have demonstrated the first 4~K silicon cavity cooled by a closed-cycle cryostat that supports a fractional frequency instability of $1\times 10^{-16}$. This is achieved by minimizing the impact of technical noise from the cryocooler. The cryostat utilizes multiple stages of thermal damping to improve the cavity's temperature instability. The mechanical connections between the cryocooler and the vacuum chamber are designed to minimize the transfer of vibrations to the cavity. \\
\indent  Further steps will be taken to optimize the system. The impact of vibrations from the cryocooler can be mitigated by improving the cavity's mounting structure, and by reducing the rigidity of the link between the cold finger and the thermal shields. Further isolation from the thermal environment is required to achieve optimal long-term performance. Feedforward can also be implemented to reduce the impact of any remaining thermal and vibrational noise~\cite{FeedforwardPRA}. By combining the technical advances of this work with AlGaAs coatings~\cite{Cole2013,Cole2016} and a longer cavity length of 20~cm, there is a clear path forward for developing a cavity with $10^{-18}$-level instability.\\
\indent The authors thank T. Asnicar, T. Brown, S. Campbell, A. Goban, R. Hutson, S. Kolkowitz, D. Leibrandt, and G.E. Marti for discussions. This work is supported by the JILA Physics Frontier Center (NSF PHY-1734006), NIST, DARPA, the Centre for Quantum Engineering and SpaceTime Research (QUEST), and Physikalisch-Technische Bundesanstalt. This project also receives funding under 15SIB03 OC18 from the EMPIR programme co-financed by the Participating States and from the European Union's Horizon 2020 research and innovation programme, and from the European Metrology Research Programme  under QESOCAS and by Q-SENSE, funded by the European Commission's H2020 MSCA RISE (69115). L.S. is supported by the NDSEG Program. E.O. acknowledges the NRC postdoctoral fellowship.
\bibliography{4K.bib} 

\end{document}